\documentclass[aps,prc,twocolumn,showpacs,nofootinbib,reprint]{revtex4-1}
\pdfoutput=1

\usepackage{blindtext}
\usepackage{amsfonts}
\usepackage{tensor}
\usepackage{graphicx}
\usepackage{pict2e}

\usepackage{hyperref}
\usepackage{makecell}
\usepackage{amssymb, amsmath,environ}

\usepackage[utf8]{inputenc}

\usepackage{mathtools}   

\DeclareMathOperator{\sech}{sech}

\DeclareMathOperator{\arccosh}{arccosh}

\DeclareMathOperator{\arctanh}{arctanh}

\usepackage{color}


\newcommand{\bse}{\begin{subequations}}
	\newcommand{\ese}{\end{subequations}}
\newcommand{\be}{\begin{equation}}
\newcommand{\ee}{\end{equation}}

\usepackage{lineno}

\newcommand{\bea}{\begin{eqnarray}}
\newcommand{\eea}{\end{eqnarray}}
\newcommand{\ba}{\begin{array}}
	\newcommand{\ea}{\end{array}}

\newcommand{\nn}{{\nonumber}}

\newcommand{\Q}{\mathcal{Q}}

\newcommand{\V}{\mathcal{V}}

\newcommand{\la}{\langle}
\newcommand{\ra}{\rangle}

\newcommand{\rms}{\text{rms}}
\newcommand{\fo}{\text{fo}}

\newcommand{\tot}{\text{tot}}

\newcommand{\mmax}{\text{max}}
\newcommand{\crit}{\text{crit}}
\newcommand{\hyd}{\text{hyd}}
\newcommand{\cor}{\text{core}}
\newcommand{\coron}{\text{corona}}
\newcommand{\qgp}{\text{QGP}}

\begin{document}
	
\title{Smallest QCD droplet and multiparticle correlations in p--p collisions}
\author{Seyed Farid Taghavi} 
\email{s.f.taghavi@tum.de}

\affiliation{Physik Department E62, Technische Universit\"{a}t M\"{u}nchen, James Franck Str.~1, 85748 Garching, Germany}

\begin{abstract}

The collective evolution of produced matter in heavy-ion collisions is effectively described by hydrodynamics from time scales greater than the inverse of the temperature, $\tau \gtrsim 1/T$. In the context of the Gubser solution, I show that the hydrodynamization condition $\tau \, T \gtrsim 1$  is translated into an allowed domain in the spatial system size and the final multiplicity for hydrodynamics applicability. It turns out that the flow measurements in p--p collisions are inside the domain of validity. I predict that by approaching the boundaries of the allowed domain the hydrodynamic response to the initial ellipticity changes its sign. I follow a rather model-independent approach for the initial state where, instead of modeling the initial energy density of individual events, the initial system size and ellipticity event-by-event fluctuation are modeled. The model, initial state fluctuation+Gubser solution+Cooper-Frye freeze-out, describes the multiplicity and transverse momentum dependence of two-point and four-point correlation functions ($c_2\{2\}$ and $c_2\{4\}$) in an accurate agreement with p--p collision experimental measurements. In particular, the sign of the four-point correlation function is the same as the observation, which failed to be described correctly in previous studies. I also predict a signal for the sign change in the hydrodynamic response that can be inspected in future experimental measurements of two-point and four-point correlation functions at lower multiplicities.

\end{abstract}

\maketitle

\section{Introduction}

In 2010, CMS collaboration revealed a peculiar observation of long-range correlation in p--p collisions \cite{Khachatryan:2010gv} which is considered as a signature of collective evolution. Later on, this observation was confirmed  by different experimental collaborations for different small systems (p--p, pAu, dAu, ${}^3$HeAu, pPb) at LHC \cite{Abelev:2014mda,Khachatryan:2016txc,Aaboud:2016yar,Aaboud:2017blb} and RHIC \cite{Adamczyk:2015xjc,PHENIX:2018lia}. Over the past years, there have been ongoing debates on the origin of the observed correlation. Efforts to explain the observed phenomena have been made from different perspectives, e.g., to describe the phenomenon via kinetic theory \cite{Kurkela:2018qeb,Kurkela:2019kip} or to link the correlation to the initial stages of the collision (for review see Ref.~\cite{Strickland:2018exs}). 

The present paper belongs to the category of studies that intend to demonstrate the observed phenomena using conventional hydrodynamics \cite{Bozek:2011if,Bozek:2013uha,Niemi:2014wta,Weller:2017tsr,Mantysaari:2017cni,Gallmeister:2018mcn,Heinz:2019dbd}. The strategy that I pursue in the present paper is the following: studying hydrodynamic evolution such that, first, it has essential features to explain the real data, and, second, it is still simple enough to monitor an event evolution anatomy clearly. To this end, the best choice in the author's opinion is the analytical solution of relativistic hydrodynamic equations for conformal fluids, the Gubser solution, and perturbation on top of that \cite{Gubser:2010ze,Gubser:2010ui}. Despite the idealized assumptions for the symmetries of the system, the Gubser solution has been used to investigate, to an extent, realistic scenarios in heavy-ion collisions. As an example, the power spectrum of heavy-ion collisions is obtained by studying the evolution of narrow peaked hot spots on top of a smooth background in Refs.~\cite{Staig:2011wj,Gorda:2014msa}. 

In the present paper, I consider the rotationally symmetric Gubser solution, which is perturbed to acquire an initial elliptical shape \cite{Gubser:2010ui}. The free parameters of the Gubser solution are carefully mapped to the more standard quantities such as the root mean square (rms) radius, ellipticity, and the total energy in the transverse direction of the system. Using this map together with the equation of state (e.o.s.), the initiation time of the evolution, the mass of the final particles, and the freeze-out temperature, one obtains an estimation for the elliptic flow of any given initial energy distribution. Considering the consistency of the Gubser solution, I find the domains of hydrodynamics applicability in terms of the spatial system size and final multiplicity. One observes that flow measurements of p--p collisions are inside the domain of validity. I also describe the measured two-particle, $c_2\{2\}$, and four-particle, $c_2\{4\}$, correlations as a function of transverse momentum and charge multiplicity. Moreover, I predict an experimental signal in multiparticle correlation functions in p--p collisions to indicate whether one approaches the boundaries of the validity domain.
Indeed, due to the several simplifications and idealizations in the Gubser solution, the present model is not applicable to gain insight into hydrodynamic transport coefficients such as shear viscosity over entropy density. A more realistic hydrodynamic simulation is essential for such studies. I justify, however, that the computations are accurate enough for the purposes I pursue in this paper. 

The paper is organized as follows: In section~\ref{secII}, I investigate the domain of applicability of hydrodynamics as a function of the initial system size and the total transverse energy. In section~\ref{secIII}, the model is introduced and compared with the outcome with {\tt iEBE-VISHNU}. The model is compared with experimental p--p collision measurements in section~\ref{secIV}, and finally one finds the conclusion in section~\ref{secV}. The computational details of the model are presented in appendix~\ref{appA}. Further comparisons between predictions of the model and {\tt iEBE-VISHNU} can be found in appendix~\ref{validationApp}.

 \section{Smallest system Size for Hydrodynamic Description}\label{secII}
 
  I start by highlighting the main aspects of the Gubser solution  \cite{Gubser:2010ze,Gubser:2010ui}. Before that, I determine the coordinate systems employed in this paper. I interchangeably use hyperbolic-cylindrical coordinates $x^\mu=(\tau,r,\phi,\eta)$ ($\tau=\sqrt{t^2-z^2}$ is proper time, $\eta=\arctanh(z/t)$ is space-time pseudorapidity, and $(r,\phi)$ is the polar coordinate in the transverse plane) and the de~Sitter coordinate $\hat{x}^\mu=(\rho,\theta,\phi,\eta)$ ($\rho$ is de~Sitter time, and $\theta$ is de~Sitter radial coordinate in the transverse direction). Two coordinates are related to each other via $q\,\tau=\sech\rho/(\cos\theta-\tanh\rho)$ and $q\,r=\sin\theta/(\cos\theta-\tanh\rho)$ where $q$ is a free parameter. 
   \begin{figure}[t]
  	\begin{center}		
  		\includegraphics[scale=0.65]{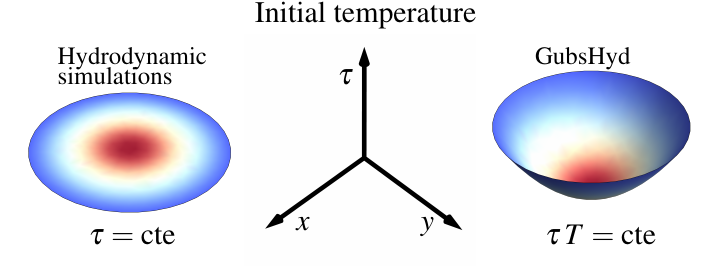}				
  		\caption{The initiation of hydrodynamic evolution at $\tau=\tau_{\text{hyd}}$ (left) and $\rho=\rho_{\text{hyd}}$ (right). } 
  		\label{InitiationEnergy}
  	\end{center}
  \end{figure}
 In the de~Sitter coordinate, the Gubser solution for the energy density of a boost-invariant ideal conformal fluid reads as 
 \bea\label{gubsersolution}
 \epsilon_b(\rho,\theta)=\frac{1}{\tau^4}\frac{\hat{\epsilon}_0}{\cosh^{8/3}\rho},
 \eea 
 where $\tau$ should be written in terms of $\theta$ and $\rho$ with the transformation mentioned earlier. In the above, $\hat{\epsilon}_0$ is a free parameter.  An elliptic perturbation on top of the rotationally symmetric background solution (i.e. Eq.~\eqref{InitiationEnergy}) leads to $\epsilon(\rho,\theta,\phi)\approx\epsilon_b(\rho,\theta)\left[1+4\lambda \,\delta_2(\rho) \,y(\theta,\phi) \right]$ with $ 	y(\theta,\phi)=-\sqrt{3/8}Y_{2,2}+Y_{2,0}/2-\sqrt{3/8}Y_{2,-2}$ ($Y_{l,m}\equiv Y_{l,m}(\theta,\phi)$ is the spherical harmonic function). The fluid velocity is given by $u_\mu=\tau(\partial \hat{x}^\nu/\partial x^\mu)\hat{u}_\nu$ where  $\hat{u}_\mu=(-1,\;\lambda\,\delta \hat{u}_\theta,\;\lambda\,\delta \hat{u}_\phi,0)$, $\lambda$ is a small free parameter,  
  and $\delta\hat{u}_i= \tilde{\delta}_2(\rho)\;\partial_i y(\theta,\phi)$ for $i=\theta,\phi$ \cite{Gubser:2010ze,Gubser:2010ui,Staig:2011wj} (see a more detailed review in appendix~\ref{appA}). The hydrodynamic equations lead to a linear system of differential equations for $\delta_2(\rho)$ and $\tilde{\delta}_2(\rho)$ which can be solved analytically for ideal hydrodynamics. The isotropic initial fluid velocity assumption at de~Sitter time $\rho=\rho_\hyd$ fixes the initial value for the system of equations as  $\delta_2(\rho_\hyd)=1$ and $\tilde{\delta}_2(\rho_\hyd)=0$. In this paper, I do not present the Navier-Stokes solution, which reveals instability as one expects.
 The problem could be cured in a causal hydrodynamics framework. However, the Gubser solution  for causal hydrodynamics has been found only in certain limits \cite{Marrochio:2013wla,Pang:2014ipa}, and perturbation on top of causal Gubser flow has not been done yet. 
 In addition to the ignored dissipative effects, the conformal symmetry prevents one from choosing a realistic e.o.s. Moreover, the hydrodynamic solution is limited to an elliptic perturbation (a linear hydrodynamic response) which also means that interaction between the ellipticity mode with itself (e.g., cubic response \cite{Noronha-Hostler:2015dbi}) and other mode-mode interactions are ignored. I address the effect of all these idealizations before I start model/data comparison.

  From now on, as an abbreviation, the Gubser hydrodynamic solution together with Cooper-Frye prescription \cite{Cooper:1974mv} for the freeze-out is called {\tt GubsHyd}. A detail of the ``model'' can be found in appendix~\ref{appA}.
  A difference between {\tt GubsHyd} and the conventional numerical hydrodynamic solutions is in their initiations. In the numerical hydrodynamic computations, the initial energy density is prepared on a $\tau=\text{cte}$ surface (Fig.~\ref{InitiationEnergy} (left)) while {\tt GubsHyd} is initiated on a surface with condition $\rho=\rho_\hyd=\text{cte}$ (Fig.~\ref{InitiationEnergy} (right)). Referring to Eq.~\eqref{gubsersolution}, one simply finds that $\rho=\text{cte}$  is equivalent to   $\tau \epsilon^{1/4}=\text{cte}$ which is subsequently equivalent to $\tau T=\text{cte}$ by using e.o.s. $\epsilon=C_0 T^4$. 
  
  It is worth reminding the reader that although I explicitly present the ideal Gubser solution in Eq.~\eqref{gubsersolution}, I implicitly assume it is only an approximation of a more general causal hydrodynamics in which the system is not in perfect local equilibrium. As a result, to apply hydrodynamics to a far-from-equilibrium initial state, a finite time is needed for hydrodynamization. In this context, the $\tau T=\text{cte}$ surface can be understood as follows: regarding studies about hydrodynamization, specifically, the computations from gauge/gravity duality, the evolution of a boost-invariant system with translational and rotational symmetry in the transverse space (Bjorken symmetry) is attracted to the hydrodynamic solutions after the time $\tau \sim 1/T$ \cite{Chesler:2009cy,Heller:2011ju}. It has been demonstrated that the hydrodynamic gradient expansion ($1/\tau T$ expansion in this context) is divergent, and its divergence is due to the presence of nonanalytic contributions $\propto e^{- z_0 \tau T}$ known as nonhydrodynamic modes   
   \cite{Heller:2013fn,Heller:2015dha,Romatschke:2017vte,Romatschke:2017acs,Heller:2016rtz}. Here, the coefficient $z_0\sim\mathcal{O}(1)$ is a positive real numerical factor depending on the underlying microscopic theory. Therefore, the nonhydrodynamic modes die out at $\tau T \sim 1$, and the hydrodynamic description works appropriately afterward.\footnote{The attractor solution has been observed in an expanding ultrarelativistic gas of hard spheres with Bjorken symmetry. It has been shown that the causal hydrodynamics can describe this system even when gradients are significant at all times of  evolution \cite{Denicol:2019lio}.}
   
   The attractor solution of a system that goes through Gubser flow has been studied in Ref.~\cite{Behtash:2017wqg} within the relativistic kinetic theory framework (see also Ref.~\cite{Chattopadhyay:2018apf}). This paper shows that hydrodynamic quantities which are parametrized  with variable $w=\tanh\rho/\tau T$ approach to an attractor at a specific value of $w$, namely $w\sim w_0$. Since $\tau T \sim \tau\epsilon^{1/4}\sim \cosh^{-2/3}\rho$, the condition $w \sim w_0$ is equivalent to the condition $\tau T\sim \text{cte}$, similar to what has been found for the systems with Bjorken symmetry. In Gubser flow, however, the temperature drops when one moves from the center to the tail of the initial energy density. As a result, it is plausible to consider that the hydrodynamization happens at different proper times depending on the transverse radius. This argument allows one to define the \textit{hydrodynamization surface}, the surface on which $\tau \, T(\tau,r)$ is constant and is in the order of unity [see Fig.~\ref{InitiationEnergy} (right)]. For a system with fewer symmetries and more complicated initial conditions, one would expect that the nonhydrodynamic modes decay at $\tau\sim 1/T(\tau,\vec{x})$, leading to a more complicated hydrodynamization surface $\tau=\tau(\vec{x})$. 
 
To compare the {\tt GubsHyd} predictions with real experimental data, one translates the  free parameters of the Gubser solution $(q,\hat{\epsilon}_0,\lambda,\rho_\hyd)$  to more standard quantities: total transverse energy $\epsilon_\tot$, rms radius $r_\rms$, and ellipticity $\epsilon_2$, 
\begin{subequations}\label{eq2}
	\begin{align}
	&\epsilon_\tot=\int r\, dr\, d\phi\; \epsilon(\tau,r,\phi), \\
	&r_\rms^2= \frac{1}{2\epsilon_\tot}\int r\, dr\, d\phi\; r^2\epsilon(\tau,r,\phi), \\
	&\epsilon_2 =-\frac{1}{2r_\rms^2\epsilon_\tot}\int r\, dr\, d\phi\; r^2\cos(2\phi)\,\epsilon(\tau,r,\phi).
	\end{align}
\end{subequations}
 The measure of these integrals reads as $\tau^2 \cosh^2\rho_\hyd\,\sin\theta\, d\theta d\phi$ in the de~Sitter coordinates. Using the measure together with the fact that the solution is initiated on the $\rho=\rho_\hyd$ surface, one obtains 
 \begin{subequations}
 	\begin{align}
&\hat{\epsilon}_0=\frac{3\,\epsilon_\tot \label{3a} \,r_\rms^2}{4\pi\cosh^{4/3}\rho_\hyd}+\mathcal{O}(\epsilon_2),\\
&\frac{1}{q^2}=r_\rms^2\left(1+3\tanh^2\rho_\hyd\right)+\mathcal{O}(\epsilon_2), \\
&\lambda=\left(\sqrt{5\pi}/3\right)\epsilon_2+\mathcal{O}(\epsilon_2^2).
 	\end{align}
 \end{subequations}
For finding $\rho_\hyd$ in terms of standard quantities, one notes that some parts of the initial energy density are about to freeze-out immediately after the initiation because the energy density of these parts is equal to the freeze-out energy density $\epsilon_\fo$. 
As a result, at the hydrodynamization surface, one has 
$\tau \epsilon^{1/4}=\tau_\hyd \epsilon_\fo^{1/4}$. Subsequently, by employing Eq.~\eqref{gubsersolution} and substituting $\hat{\epsilon}_0$ in terms of standard quantities (Eq.~\eqref{3a}), one finds
\bea\label{rho0}
\rho_\hyd=-\arccosh\left[\left(\frac{3\,r_\rms^2\, \epsilon_\tot}{4\pi\,\tau_\hyd^4 \,\epsilon_\fo}\right)^{1/4}\right]+\mathcal{O}(\epsilon_2).
\eea

\begin{figure}[t]
	\begin{center}		
		\includegraphics[scale=0.65]{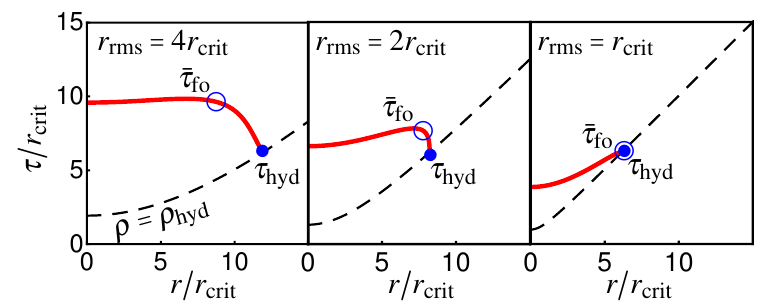}				
		\caption{Freeze-out surfaces for systems of three different sizes where $r_\text{crit}=0.1\;$fm, $\tau_\text{hyd}=0.62\;$fm/c, $C_0=11$.} 
		\label{freezeSurfTaki}
	\end{center}
\end{figure}

I employ Cooper-Frye prescription \cite{Cooper:1974mv} with Boltzmann equilibrium distribution to obtain the associated final particle distribution function:
\bea\label{cff}
\frac{dN}{d\boldsymbol{p}}=-\frac{g}{(2\pi)^3}\int p^\mu d\Sigma_\mu\;\exp\left[p_\mu u^\mu/T_\fo\right],
\eea
where $d\boldsymbol{p}\equiv d^3p/E$ and $\Sigma^\mu=(\rho,\theta_\fo(\rho,\phi),\phi,\eta)$ indicates the freeze-out surface specified  by equation $\epsilon(\rho,\theta_\fo,\phi)=\epsilon_\fo$. In Eq.~\eqref{cff}, $g$ is the degeneracy constant and $T_\fo$ is the freeze-out temperature related to $\epsilon_\fo$ via e.o.s. The equation $\epsilon(\rho,\theta_\fo,\phi)=\epsilon_\fo$ for the unperturbed Gubser solution (Eq.~\eqref{gubsersolution}) leads to the following equation for the freeze-out surface:
\begin{equation}\label{freezeSurface}
\begin{split}
&\cos\theta_\fo(\rho)=\tanh\rho+\frac{1}{q}\left(\frac{\epsilon_\fo}{\hat{\epsilon_0}}\right)^{1/4}\sech^{1/3}\rho.
\end{split}
\end{equation}
Red curves show the freeze-out surfaces in Fig.~\ref{freezeSurfTaki} for systems with three different sizes.

Initiating the evolution on the hydrodynamization surface leads to an interesting conclusion. One defines $\bar{\tau}_\fo$ as the time when the last fluid cell of the system is frozen out (blue open circles in Fig.~\ref{freezeSurfTaki}). On the other hand, by definition, $\tau_\hyd$ on the hydrodynamization surface is also a member of the freeze-out surface (blue bullets show $\tau_\hyd$ location in Fig.~\ref{freezeSurfTaki}). One can interpret $\tau_\hyd$ as the time at which the last fluid cell is hydrodynamized. One notes that the condition $\tau_\hyd\leq \bar{\tau}_\fo$ should always be satisfied; otherwise, the last frozen-out cell has not enough time to be hydrodynamized. It turns out that there are some values for $(r_\rms,\epsilon_\tot,\epsilon_\fo,\tau_\hyd)$ that correspond to no valid hydrodynamic solution for making a hydrodynamization surface. With that, one concludes a lower bound on the system size. 
 Considering that the argument in Eq.~\eqref{rho0} should be greater than unity to have a real-valued  $\rho_\hyd$, one obtains the following lower bound on the system size, 
\bea\label{theInequalityI}
r_\rms \geq r_\crit,\qquad r_\crit=\sqrt{\frac{4\pi}{3}}\left[ \tau_\hyd^2 \,\epsilon_\fo^{1/2}\right] \epsilon_\tot^{-1/2}.
\eea
In Fig.~\ref{freezeSurfTaki}, the freeze-out surface for a system at a critical size is depicted in the right panel. 

The above arguments can be considered as a hint for the presence of the same lower bound for any nonideal/nonconformal hydrodynamic systems. Na\"ive reasoning is as follows: Consider a nonequilibrium system with a fixed amount of total transverse energy $\epsilon_\tot$. The smaller the system size, the sharper the concentration of energy. Compared to a system with a larger size, a system with more concentration of energy would expand faster because of the more considerable ``pressure" gradient. Therefore, one might consider cases in which the system expands so fast that the nonhydrodynamic modes do not have enough time to decay before the nonequilibrium system reaches the phase transition boundary. For such a system, equation $\tau \epsilon^{1/4}(\tau,\vec{x})=\tau_\hyd \epsilon_\fo^{1/4}$ has no solution, and the hydrodynamization surface $\tau(\vec{x})$ does not exist. For more robust evidence, computations beyond the hydrodynamic regime are needed. In fact, the bound \eqref{theInequalityI} is compatible with that obtained from studying two colliding shock waves in the context of numerical holography \cite{Chesler:2015bba,Chesler:2016ceu}. One can estimate the averaged initial energy density as $\bar{\epsilon}_\text{init}\sim \epsilon_\tot/\pi r_\crit^2$. Using $T_\text{eff}=(4\bar{\epsilon}_\text{init}/3\pi^4)^{1/4}$ (Eq.~(7) in Ref.~\cite{Chesler:2015bba}), the condition $r_\crit\,T_\text{eff} \sim 1$ (Eq.~(10) in Ref.~\cite{Chesler:2015bba}) is equivalent to $r_\crit\,\epsilon_{\text{tot}}^{1/2} \sim 1$ in Eq.~\eqref{theInequalityI}. The other possibility is using the kinetic theory with nontrivial energy distribution in the transverse direction, such as that studied in Refs.~\cite{Kurkela:2018qeb,Kurkela:2019kip}. Studying the hydrodynamic attractors for such a system would lead to evidence for the presence of a lower bound on the system size at which hydrodynamics is applicable. 

From Eq.~\eqref{freezeSurface}, one can also find the de~Sitter time when the last fluid cell is frozen out via $\cos\theta_\fo(\rho_\mmax)=1$. Consequently, the condition $\rho_\hyd=\rho_\mmax$ means there is no ``time'' left for hydrodynamic evolution. With that, one concludes an upper bound,
\bea\label{upperbound}
r_\rms \leq R_\crit(\epsilon_\tot),
\eea
 for the system size corresponding to the case that there is not enough energy density deposited into the given region to produce a deconfined matter. Translating variables into standard quantities, one can solve the equation $\cos\theta_\fo(\rho_\hyd)=1$ in Eq.~\eqref{freezeSurface} to find $R_\crit(\epsilon_\tot)$ in Eq.~\eqref{upperbound}.

\section{Hydrodynamic response to initial ellipticity}\label{secIII}

\begin{figure}[t]
	\begin{center}		
		\includegraphics[scale=0.60]{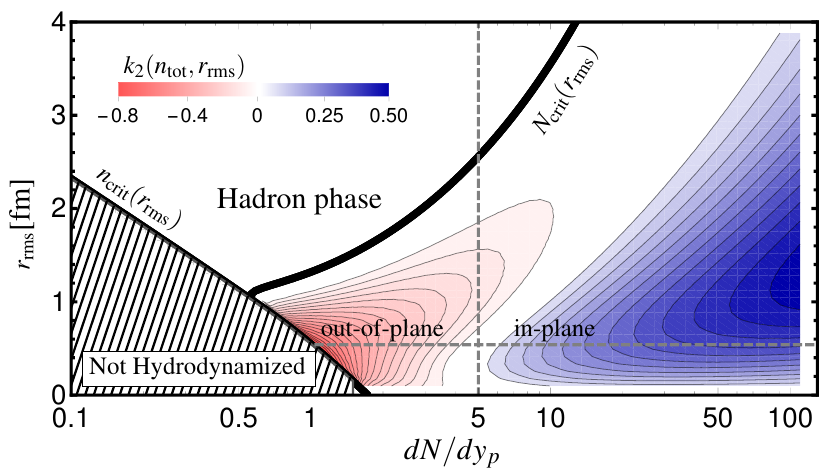}				
		\caption{Domain of hydrodynamics validity in the $(n_\tot,r_\rms)$ phase space. The {\tt GubsHyd} output for $k_2(n_\tot,r_\rms)$ ($0.3<p_T<3.0\,\text{GeV}$) is shown as a contour plot ($n_{\text{tot}}\equiv dN/dy_p$).  } 
		\label{k2}
	\end{center}
\end{figure}

The domain of hydrodynamic validity has been investigated in the previous section. To determine whether the experimental measurements for p--p collision are inside this domain, one needs to relate the hydrodynamic solutions to the final particle distribution. Having found the freeze-out surface, one can compute the final particle distribution by performing the integrals in the Cooper-Frye formula. The details of the computation are presented in appendix~\ref{appA}. Furthermore, these computations are implemented in a Mathematica package available as an ancillary file. In the following, I briefly mention some of the functions discussed in appendix~\ref{appA}: the total multiplicity in the unit rapidity $n_{\tot}(r_\rms,\epsilon_\tot)$ ($n_{\text{tot}}\equiv dN/dy_p$); the translation of critical values $r_\crit(\epsilon_\tot)$ and $R_\crit(\epsilon_\tot)$ in terms of multiplicity in unit rapidity, $n_\crit(r_\rms)$ and $N_\crit(r_\rms)$; 
the particle distributions in unit space-time pseudorapidity $dN/p_Tdp_Td\eta$ and $dN/d\eta$; and finally and most importantly $k_2(n_\tot,r_\rms)$ and $k_2(\epsilon_\tot,r_\rms)$ (and their $p_T$ dependent versions) which relate the initial ellipticity $\epsilon_2$ to the elliptic flow, the second Fourier coefficient of the final particle distribution in the azimuthal direction,
\bea\label{linearEq}
v_2(n_\tot,r_\rms)=k_2(n_\tot,r_\rms)\,\epsilon_2.
\eea 
The initial hydrodynamization time $\tau_\hyd$, the final particle mass $m$, the freeze-out energy density $\epsilon_\fo$, and the  constant $C_0$ in e.o.s. are considered to be fixed in the model.

In Fig.~\ref{k2}, the contour plot of $k_2(n_\tot,r_\rms)$ for pions $m\approx136\;$MeV is depicted for $0.3<p_T<3\,\text{GeV/c}$. Here, I have fixed $\tau_\hyd=0.9\,\text{fm/c}$,  $\epsilon_\fo=0.18\;\text{GeV/fm}^3$, and $C_0=13$. 
The latter is chosen close to the plateau in s95p-v1 \cite{Huovinen:2009yb} at temperatures above the critical point where most of the evolution period is in this range. The functions $n_\crit(r_\rms)$ and $N_\crit(r_\rms)$ are also shown with thick black curves. What one sees from the figure is that the system is hydrodynamized in the range $n_\tot>n_\crit(r_\rms)$.
 Moreover, in the valid domain of hydrodynamics, there are two different regions: the \textit{in-plane response} region where particles are mostly emitted along the minor axis of the initial elliptic shape (blue contours) and the \textit{out-of-plane response} regions where most of the particles are emitted along the major axis (red contours). As it will be discussed later, the signature of in-plane/out-of-plane transition is traceable in multiparticle correlations in p--p collisions. 

Due to the idealizations that have been made, one might be doubtful about the applicability of the presented hydrodynamic model in explaining the experimental data. To show that {\tt GubsHyd} is accurate enough for the purpose I pursue in this paper, I compare a causal hydrodynamic simulation, {\tt  iEBE-VISHNU} ({\tt MC-Glauber}+{\tt VISH2+1})  \cite{Shen:2014vra} with {\tt GubsHyd}. Since the interested quantities are those which are averaged over an ensemble of events, I show that all the corrections can be encapsulated in a generic constant $\chi$,
\bea\label{scaleLinear}
v_2=\chi\, k_2(\epsilon_\tot,r_\rms)\,\epsilon_2.
\eea
It will be shown that $\chi$ can be absorbed into the ellipticity fluctuation width, a free parameter in the model, and has no impact on the predictions in the present paper.

\begin{figure}
	\begin{center}
		\hspace*{-0.4cm}\includegraphics[scale=0.65]{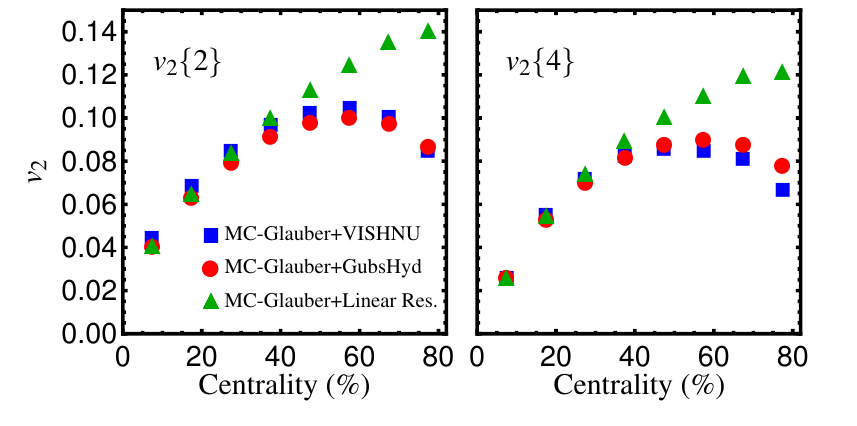}
		\caption{Two- and four-particle correlations ($v_2\{2\}$ and $v_2\{4\}$) from {\tt MC-Glauber}+{\tt VISH2+1} and {\tt MC-Glauber}+{\tt GubsHyd}.} 
		\label{GubsHydVISHNUcomp}
	\end{center}
\end{figure}

I employ {\tt iEBE-VISHNU} with {\tt MC-Glauber} as an initial state model for Pb--Pb collisions at $\sqrt{s_{NN}}=2.76\,\text{TeV}$. The simulation is based on solving 2+1 boost-invariant causal hydrodynamic equations with a fixed $\eta/s=0.08$. The initial time of hydrodynamic evolution is  $\tau=0.6\;\text{fm/c}$, and freeze-out energy density is fixed to  $\epsilon_\fo=0.18\;\text{GeV/fm}^3$. The afterburner is not included in this simulation. The events are classified into 16 bins  between $0$ and $80\%$ centrality classes, and at each bin 14\,000 events are generated. The elliptic flow is computed for charged pions with the momentum in the range $0.28<p_T<4\,\text{GeV}$ at each event. After that, the probability density function (p.d.f.) of elliptic flow fluctuation $p_v(v_2)$ at each centrality class is obtained, and eventually, I compute the two first cumulants of the p.d.f. \cite{Borghini:2001vi}
\begin{subequations}\label{v2}
	\begin{align}
	c_2\{2\}&=\la v_2^2 \ra_v\label{v2a}\qquad\qquad\equiv (v_2\{2\})^2,\\
	c_2\{4\}&=\la v_2^4\ra_v-2\la v_2^2\ra_v^2\label{v2b}\, \equiv -(v_2\{4\})^4,
	\end{align}
\end{subequations}
where $\la \cdots \ra_v$ stands for averaging with respect to $p_v(v_2)$. 
Now, from the same set of initial states, I compute $(\epsilon_\tot,r_\rms,\epsilon_2)$ at each event. Setting $\tau_\hyd=0.6\,\text{fm/c}$, $m=136\;\text{MeV}$,  $\epsilon_\fo=0.18\;\text{GeV/fm}^3$, and $C_0=13$, I employ {\tt GubsHyd} (the function in the second line of Eq.~\eqref{50}) for $0.28<p_T<4\,\text{GeV}$ and use $v_2= k_2(\epsilon_\tot,r_\rms)\,\epsilon_2$ to obtain $v_2$ at each event. Eventually, one has 14\,000 events for each centrality class where its elliptic flow is evaluated from {\tt GubsHyd}. Similar to {\tt iEBE-VISHNU} events, one can compute $v_2^\text{\tiny IdealGub}\{2\}$ and $v_2^\text{\tiny IdealGub}\{4\}$ from {\tt GubsHyd} as well. One notes from Eq.~ \eqref{v2} that the constant $\chi$ in Eq.~\eqref{scaleLinear} corrects the cumulants as
\bea\label{37}
v_2^{\text{\tiny Gub}}\{2k\}=\chi\, v^{\text{\tiny IdealGub}}_2\{2k\}.
\eea
The results are shown by red filled circles in Fig.~\ref{GubsHydVISHNUcomp} where $\chi=0.39$ is chosen. As seen from the figure, there is a good agreement between {\tt iEBE-VISHNU} and {\tt GubsHyd} in the given range of centralities. 

Eq.~\eqref{scaleLinear} is different from that mentioned in  Ref.~\cite{Gardim:2011xv} as a linear hydrodynamic response:
\bea\label{54}
v_2\approx \tilde{k}_2 \;\epsilon_2,
\eea
where $\tilde{k}_2$ is a constant unlike $k_2(\epsilon_\tot,r_\rms)$. 
To show that the observed agreement is not trivially inherited from the initial state, one can obtain the same cumulants as Eq.~\eqref{37} by employing Eq.~\eqref{54}:
\bea\label{39}
v_2^{\text{\tiny Glaub}}\{2k\}=\tilde{k}_2 \,\epsilon_2\{2k\}.
\eea
In the above, $\epsilon_2\{2k\}$ can be obtained from Eq.~\eqref{v2} by replacing the variable $v_2$ with $\epsilon_2$ while the average $\la \cdots \ra_\epsilon$ is performed using the distribution $p_\epsilon^{\text{\tiny Glaub}}(\epsilon)$ associated with the ellipticity fluctuation in the {\tt MC-Glauber} model.
 In Fig.~\ref{GubsHydVISHNUcomp}, $v_2^\text{\tiny Glaub}\{2\}$ and $v_2^\text{\tiny Glaub}\{4\}$ with  $\tilde{k}_2=0.25$ are plotted by the green filled triangles.
Although linear response explains the causal hydrodynamic simulations in central collisions, it fails to explain it in peripheral collisions.

The above investigation indicates that, concerning many event averages, the effect of idealizations in the {\tt GubsHyd} model can be corrected approximately via a constant value $\chi$. The numerical value of $\chi$ is irrelevant in the analysis I perform in the next section. For completeness, the {\tt GubsHyd} predictions  have been compared with {\tt iEBE-VISHNU} for $p_T$ spectrum, differential $v_2\{2\}$, and $v_2\{4\}$; and centrality dependence of the multiplicity for charged pions in appendix~\ref{validationApp}.

 \section{Two- and Four-particle correlations in $\text{p}$--$\text{p}$ collisions}\label{secIV}

The initial state is the final piece in the model presented in this paper. In this paper, I follow a rather model-independent approach for the initial state. Instead of generating an ensemble of initial energy densities, $\epsilon_{2}$ and $r_\rms$ \textit{event-by-event fluctuations} are modeled. By comparing the model with data, one finds the properties of the $\epsilon_{2}$ and $r_\rms$ fluctuations. In the end, one can check different initial state models (or one model with different parameter tuning) to determine which one can reveal the same fluctuating properties.

Assuming $r_\rms$ and $\epsilon_2$ fluctuate  event by event independently, the p.d.f. of the two-dimensional (2D) variable $(\epsilon_2,r_\rms)$ can be written as a product of ellipticity distribution $p_\epsilon(\epsilon_2)$ and rms radius distribution $p_r(r_\rms)$;
 $$p_{\text{init}}(\epsilon_2,r_\rms)=p_\epsilon(\epsilon_2)\,p_r(r_\rms).$$ To find the  p.d.f. of  elliptic flow, one changes the variable $(\epsilon_2,r_\rms)$ to $(v_2,r_\rms)$ via Eq.~\eqref{linearEq}. The new p.d.f. reads as $p_{\epsilon}(v_2/k_2^{\text{caus}})p_r(r_\rms)/k_2^{\text{caus}}$ where 
 $$k^{\text{caus}}_2\equiv \chi k_2(n_\tot,r_\rms).$$ 
 By averaging out the variable $r_\rms$, a p.d.f. for $v_2$ fluctuation is obtained,
   \begin{equation} \label{FlowModel}
 \begin{split}
 p_v(v_2;n_\tot)=\int \frac{dr_\rms}{k_2^{\text{caus}}} p_\epsilon(v_2/k_2^{\text{caus}})\,p_r(r_\rms),
 \end{split}
 \end{equation}
 where $n_\tot$ in the argument appears as a parameter. 
 
   Now, I estimate each $p_\epsilon(\epsilon_2)$ and $p_r(r_\rms)$ separately. The ellipticity distribution can be expanded as the following series \cite{Abbasi:2017ajp,Mehrabpour:2018kjs}:
 \bea\label{pepsilon}
 p_\epsilon(\epsilon_2)=\frac{\epsilon_2}{\sigma_\epsilon^2}e^{-\epsilon_2^2/2\sigma_\epsilon^2}[1+\frac{\Gamma_2^\epsilon}{2} L_2(\epsilon_2^2/2\sigma_\epsilon^2)+\cdots],
  \eea
 where $L_2(\epsilon_2^2/2\sigma_\epsilon^2)$ is the second Laguerre polynomial and 
 \bea\label{kurtosis}
 \Gamma_2^\epsilon\equiv-(\epsilon_2\{4\}/\epsilon_2\{2\})^4
 \eea 
  is the kurtosis of the distribution.  One could use the power distribution introduced in Ref.~\cite{Yan:2013laa} for the ellipticity fluctuation in small systems. Compared to power distribution which is obtained for point-like sources, the distribution in Eq.~\eqref{pepsilon} is more general.
   It turns out that the ellipsis in Eq.~\eqref{pepsilon} expansion is irrelevant to $v_2\{2\}$ and $v_2\{4\}$ quantities. Therefore, concerning the ellipticity fluctuation, no physical assumption is imposed here.  
  For $r_\rms$, I assume a Gaussian distribution
  \bea\label{pr}
  p_r(r_\rms)=\frac{r_\rms}{\sigma_r^2}e^{-r_\rms^2/2\sigma_r^2},
  \eea
  with free parameter $\sigma_r$ and ignore other corrections to it.
  
  By substituting Eq.~\eqref{pepsilon} into Eq.~\eqref{FlowModel} and using relations in Eqs.~\eqref{v2} and $k_2^{\text{caus}}\approx \chi\,k_2$, one finally obtains the model predictions:
\begin{subequations}\label{c2s}
	\begin{align}
	v_2\{2\}&=\chi\,\sigma_\epsilon\,\sqrt{2\la k_2^2 \ra_r},\label{9a}\\
	v_2\{4\}&=\chi\,\sigma_\epsilon\,\left[8\la k_2^2 \ra_r^2-4(2+\Gamma_2^\epsilon)\la k_2^4 \ra_r\right]^{1/4},\label{9b}
	\end{align}
\end{subequations}
with free parameters $\sigma_\epsilon$, $\Gamma_2^\epsilon$, and $\sigma_r$.  In the above,  $\la \cdots \ra_r$ refers to averaging with respect to $p_r(r_\rms)$.
One obtains the $p_T$ dependent or $p_T$ integrated predictions for $v_2\{2\}$ and $v_2\{4\}$ from {\tt GubsHyd} by substituting the $p_T$ dependent or $p_T$ integrated $k_2$ function into relations \eqref{c2s} (see Eq.~\eqref{a38}). 
 
\begin{figure}
	\begin{center}
		\includegraphics[scale=0.65]{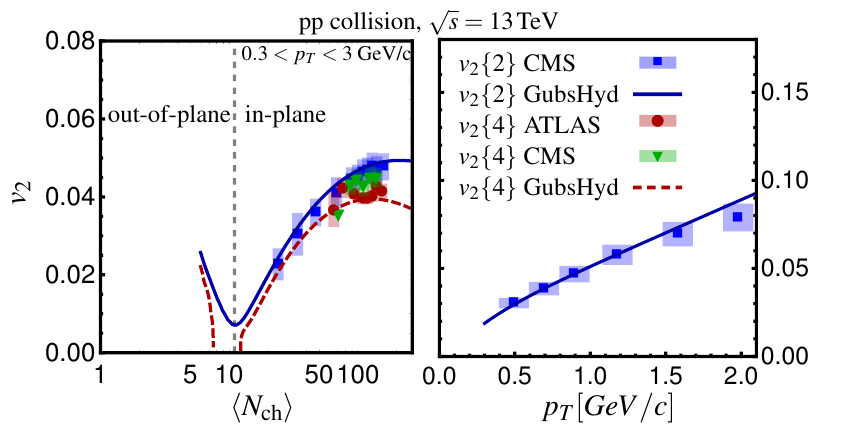}
		\caption{Comparing {\tt GubsHyd} predictions with CMS and ATLAS measurements published in Refs.~\cite{Khachatryan:2016txc,Aaboud:2017blb}.} 
		\label{ATLAS}
	\end{center}
\end{figure}

 After adding the final piece into the model, one can compare it with experimental data. In Fig.~\ref{ATLAS}, the measured $v_2\{2\}$ and $v_2\{4\}$  by CMS \cite{Khachatryan:2016txc} and ATLAS \cite{Aaboud:2017blb} collaborations
 for p--p collisions at $\sqrt{s}=13\;\text{TeV}$ are presented.
 The {\tt GubsHyd} predictions for $v_2\{2\}$ (blue curve) and $v_2\{4\}$ (red dashed curve)  are also shown where I have used the following parameters,
\bea\label{initialFromData}
\chi\,\sigma_\epsilon\approx 0.097,\quad \Gamma_2^\epsilon \approx -0.75,\quad \sigma_r\approx0.4\,[\text{fm}],
\eea
in Eq.~\eqref{c2s}. To compare with experimental data, the parameter $n_\tot$ is translated to the average charge multiplicity $\la N_\text{ch} \ra$ in the range $|\eta| <2.4 $ via
\begin{equation}
\begin{aligned}
\la N_{\text{ch}} \ra &= \frac{2}{3}\int dr_\rms \,  p_r(r_\rms) \\
&\times N_\tot(n_\tot,r_\rms,p_{T,\text{min}},p_{T,\text{max}},\eta_{\text{min}},\eta_{\text{max}})
\end{aligned}
\end{equation}
where $N_\tot$ can be found in Eq.~\eqref{A22}. In  measuring $v_2\{4\}$ in the ATLAS collaboration, the three-subevent method is employed to reduce the non-flow effect \cite{Aaboud:2017blb}. For that reason, I compare the model prediction for $v_2\{4\}$  only with ATLAS results. I did not present multiplicity dependence of $v_2\{2\}$ from the ATLAS collaboration (with peripheral subtraction method) because it follows the same trend as the CMS result (with $\eta$-gap $\Delta\eta>2$) and makes the plot hard to read.\footnote{The data points of ATLAS measurements for $v_2\{2\}$ are not available online. For the comparison with CMS data, I extracted the ATLAS measurement data points directly from the published figure. This is a second reason I have not used the ATLAS data in the figure.} Given that the idealized model {\tt GubsHyd} has been employed to extract the initial state fluctuation properties, the numbers in Eq.~\eqref{initialFromData} should be assumed as an approximate estimation.

By comparing the model with experimental data, one could extract the parameters $\Gamma_2^\epsilon$ and $\sigma_r$. However, due to the lack of knowledge about constant $\chi$, the parameter $\sigma_\epsilon$ remains undetermined. 
The term  $\chi\,\sigma_\epsilon$ is an overall factor in Eq.~\eqref{c2s} and has no impact on multiplicity and $p_T$ dependence of $v_2\{2\}$ and $v_2\{4\}$. It means irrespective to the value of $\chi$ the agreement between {\tt GubsHyd} prediction and experimental data for $v_2\{2\}$ and $v_2\{4\}$ is a triumph for the model. Moreover, an interesting physics prediction from the model can be tested by pushing the experimental measurements close to the critical multiplicity in the future. In Fig.~\ref{ATLAS}~(left), the model predicts a valley around $\la N_\text{ch} \ra=10$ shown by a vertical dashed line. This is an indication of in-plane/out-of-plane transition.  By translating $\la N_\text{ch}\ra$ to $n_\tot$, a similar vertical dashed line is shown in Fig.~\ref{k2} and a horizontal dashed line that locates $\la r_\rms\ra_r$. The latter is shown as an indicator for the average value of the system size. Moving from right to left along the horizontal dashed line in Fig.~\ref{k2}, $k_2(n_\tot,r_\rms)$ changes sign from positive to negative. Since there are only even powers of $k_2$ in Eq.~\eqref{c2s}, the sign change makes a valley in $v_2\{2\}$ and $v_2\{4\}$ versus multiplicity. Indeed, nonflow effects are more pronounced at lower multiplicities, and observing the flow signals at lower multiplicity would be experimentally challenging.
The discontinuity in $v_2\{4\}$ prediction is due to the fact that in a region of multiplicity, the terms in the bracket in Eq.~\eqref{9b} turn to negative values. One notes that the predicted valley in multiplicity dependence of $v_2\{2\}$ and $v_2\{4\}$ is based on an idealized hydrodynamic model. A framework beyond hydrodynamics is needed for a more accurate prediction.

 As a final remark, I comment on 
 the  sign of $c_2\{4\}=-(v_2\{4\})^2 $ when one initiates the energy density with a realistic model. This model must contain nucleon substructure to produce enough fluctuation.  To this end, the {\tt AMPT} event generator \cite{Lin:2004en} and {\tt T$_\text{R}$ENTo} with nucleon substructure \cite{Moreland:2014oya,Moreland:2018gsh} are examined as initial state models. 
 
 I follow Ref.~\cite{Xu:2016hmp} and employ the {\tt AMPT} event generator in the string melting mode to generate 10\,000 p--p collision initial state events at $\sqrt{s}=13\,\text{TeV}$. I map all the initiated partons inside the range $\eta<1$ into the transverse plane at each event. After that, I smear the partons with a 2D Gaussian distribution with width $\sigma_g$ weighted by the energy of each parton. After calculating $r_\rms$ and $\epsilon_2$ at each event, I compute $\sigma_r^{(\text{AMPT})}$ and $\Gamma_2^{\epsilon(\text{AMPT})}$. The results are tabulated in Table~\ref{tableKurt} for different choices of $\sigma_g$. 
\begin{table}[t!]
	\centering
	\setlength{\arrayrulewidth}{0.4pt}
	\renewcommand{\arraystretch}{1.7}
	\begin{tabular}[t]{p{1.8cm}  p{2.0cm} p{2.0cm} p{2.0cm} }
		\hline\hline
		$\sigma_g$ [fm] & $\sigma_r^{\text{AMPT}}$ [fm]  & $\Gamma_2^{\epsilon(\text{AMPT})}$ & $\Gamma_2^{v(\text{AMPT+Gubs})}$  \\
		\hline
		 0.5 & 0.48   & 0.53  & 0.80  \\
		 0.4 & 0.41   & 0.18  & 0.53  \\
		 0.3 & 0.35   & -0.17  & 0.26  \\
 		 0.2 & 0.30   & -0.48  & 0.01  \\
		 0.1 & 0.26   & -0.73  & -0.20  \\
	\hline\hline		
	\end{tabular}
	\caption{ The rms radius fluctuation width and the kurtosis of ellipticity fluctuation from {\tt AMPT}, and $\Gamma_2^{v}$ from {\tt AMPT}+{\tt GubsHyd}.}\label{tableKurt}
\end{table}%
In the fourth column, I have presented the kurtosis of elliptic flow fluctuation, 
\bea
\Gamma_2^v=c_2\{4\}/c_2^2\{2\},
\eea
from {\tt AMPT}+{\tt GubsHyd} at $\la N_\text{ch} \ra\approx150$. I compute this quantity (which is Eq.~\eqref{kurtosis} for $v_2$ fluctuation) because  the term $\chi \sigma_\epsilon$ is canceled out from the numerator and denominator, and at the same time its sign is the same as $c_2\{4\}$. A similar quantity called the effective number of sources, $N_s$, has been measured for p--p collisions by ATLAS \cite{Aaboud:2017blb} and can be translated to kurtosis by $ \Gamma_2^v=-4/(N_s+3)$. At charge multiplicity $\la N_\text{ch} \ra\approx150$, one finds the following range for the ATLAS measurement:
\bea
-0.39\lesssim\Gamma_2^{v(\text{ATLAS})}\lesssim-0.31.
\eea 
 Referring to Table~\ref{tableKurt}, one observes that the negative sign for kurtosis is more compatible with $\sigma_g= 0.1\,\text{[fm]}$. A p--p collision hydrodynamic study uses the {\tt HIJING}+{\tt iEBE-VISHNU} model in which a positive sign for $c_2\{4\}$ has been obtained \cite{Zhao:2017rgg}. In addition to examining two values $\sigma_g =0.2,0.4\,[\text{fm}]$ for the smearing width, a range of other initial state and hydrodynamic parameters is tested in this paper. For all cases, the simulation outcome for $c_2\{4\}$ (and consequently $\Gamma_2^v$) has a positive sign. The {\tt GubsHyd} model with the parameters mentioned in Eq.~\eqref{initialFromData} returns the value $\Gamma_2^{v(\text{Gubs})}=-0.41$ which is the most accurate result compared to {\tt AMPT}+{\tt GubsHyd} and {\tt HIJING}+{\tt iEBE-VISHNU}.

By comparing the content of Table~\ref{tableKurt} with Eq.~\eqref{initialFromData}, one observes tension between {\tt AMPT} initial state fluctuation properties with those extracted from the measurement directly by employing {\tt GubsHyd} (Eqs.~\eqref{c2s}). The rms radius fluctuation is more compatible with smearing width $\sigma_g = 0.4\,[\text{fm}]$ while kurtosis is more compatible with $\sigma_g = 0.1\,[\text{fm}]$.

The initial state model {\tt T$_\text{R}$ENTo} with nucleon substructure  successfully describes the initial state of large and small collision systems. I use this model with parameters  calibrated by Bayesian analysis \cite{Moreland:2018gsh}. For the calibration, experimental measurements of Pb--Pb and p--Pb collisions at $\sqrt{s_\text{NN}}=5.02\;$TeV have been used. A full description of the parameters can be found in Ref.~\cite{Moreland:2018gsh}. Here, I focus on two parameters that impact the initial state fluctuation: $n_c$, the number of constituents inside the nucleon, and $\chi_\text{struct}$ which controls the width of the constituents ($\chi_\text{struct}=0$ corresponds to a nucleon with well-separated hot spots and $\chi_\text{struct}=1$ corresponds to a nucleon as a single Gaussian blob). The calibrated values (maximum a \textit{posteriori}) of these parameters are $n_c=6.3$ and $\chi_\text{struct}=0.35$. I generate 100\,000 p--p collision initial state events and compute kurtosis of ellipticity fluctuation and rms radius fluctuation width:
\bea
\Gamma_2^{\epsilon(\tt\text{T$_\text{R}$EN})}\approx 0.09,\qquad \sigma_r^{\tt\text{T$_\text{R}$EN}}\approx 0.47\,\text{[fm]}.
\eea
The result shows tension between {\tt T$_\text{R}$ENTo} initial state fluctuation and Eq.~\eqref{initialFromData}. I also compute the hydrodynamic response to the {\tt T$_\text{R}$ENTo} initial state by using {\tt GubsHyd}. At $\la N_\text{ch} \ra\approx150$, {\tt T$_\text{R}$ENTo}+{\tt GubsHyd} leads to $\Gamma_2^{v(\tt\text{T$_\text{R}$EN}+Gubs)}=0.26$, meaning $c_2\{4\}$ is positive. I checked all combinations of $n_c=2,4,6,8$ and $\chi_\text{struct}=0.1,0.3,0.5,0.7$. The largest negative kurtosis for $\epsilon_2$ fluctuation is obtained by $n_c=8$ and $\chi_\text{struct}=0.1$ as $\Gamma_2^{\epsilon(\tt\text{T$_\text{R}$EN})}\approx -0.18$. For rms radius fluctuation width, one finds $\sigma_r^{\text{T$_\text{R}$EN}}\approx 0.37\,$[fm] in this choice of parameters. Even in this case, the kurtosis of elliptic flow fluctuation is positive, $\Gamma_2^{v(\tt\text{T$_\text{R}$EN}+Gubs)}\approx 0.18$.

A more realistic hydrodynamic simulation that includes more stages (namely, pre-equilibrium and afterburner) is needed for a rigorous conclusion. Given that the estimations based on {\tt AMPT} and {\tt T$_\text{R}$ENTo} could not predict a correct sign for $c_2\{4\}$ irrespective of its actual value, one might deduce that there is a missing piece in modeling the initial states or in modeling the collective evolution in p--p collisions. I will elaborate on this conclusion in the next section.

\section{Conclusion}\label{secV}

In this paper, I introduced {\tt GubsHyd}, a simple hydrodynamic ``model" based on the Gubser solution.  I showed that there is a lower bound for the size of a hydrodynamized system. I found that the flow measurements of p--p collisions are inside the domain of validity. I checked the validity of the presented simple model by comparing  {\tt MC-Glauber}+{\tt GubsHyd} and {\tt MC-Glauber}+{\tt VISH2+1} predictions in Pb--Pb collisions. I discussed the limitations of {\tt GubsHyd} that should be taken into account in the conclusions.  I found that three normalization/moments of the initial energy density, namely total energy density in the transverse direction, rms radius, and ellipticity (see Eqs.~\eqref{eq2}), approximate the initial energy density and lead to a reasonable agreement between {\tt GubsHyd} and {\tt iEBE-VISHNU} for elliptic flow predictions in a wide range of centrality classes.
The model could explain $v_2\{2\}$ and $v_2\{4\}$ measured by CMS and ATLAS collaborations for high multiplicity p--p collisions. Moreover, for  systems close to the smallest QCD droplet, the model predicts an experimental signal in $v_2\{2\}$ and $v_2\{4\}$ measurements at lower multiplicities where in-plane/out-of-plane hydrodynamic response transition happens.

Employing the simplicity of the Gubser solution, I found a smooth function for a hydrodynamic response to the initial ellipticity. This response function is given in terms of final state multiplicity and rms radius of the initial state $k_2(n_\tot,r_\rms)$ (see Fig.~\ref{k2}). Given that $\epsilon_2$ and $r_\rms$ play the most significant roles in the final value of elliptic flow, I modeled ellipticity and rms radius fluctuations. Then, using the smooth function $k_2(n_\tot,r_\rms)$, I explicitly obtained elliptic flow fluctuation in terms of initial state fluctuation properties (Eqs.~\eqref{c2s}). Since I do not use any specific model for the initial state energy density, I could extract the fluctuation properties from the experimental measurements.
By comparing {\tt GubsHyd} predictions for two- and four-particle correlations with those measured for p--p collisions, I found that the kurtosis of ellipticity fluctuation should be  -3/4 approximately (see Eq.~\eqref{initialFromData}). I examined two initial state models, {\tt AMPT} and {\tt T$_\text{R}$ENTo}, in a wide range of parameters. None of these models could reproduce such a significant negative value for the kurtosis and, at the same time, a reasonable value for rms radius fluctuation width. This observation could explain the unsuccessful hydrodynamic predictions for $c_2\{4\}$ in p--p collisions in recent studies.  

Regarding the sign of $c_2\{4\}$, a more realistic hydrodynamic simulation is essential for the future. However, considering the validation of {\tt GubsHyd} with {\tt iEBE-VISHNU} in Sec.~\ref{secIII} and appendix~\ref{validationApp}, I would expect that the conclusions based on {\tt GubsHyd} should be close to a conventional causal hydrodynamic study. One notes that the effect of hydrodynamization surface becomes important at lower multiplicities. The four-particle correlation, $c_2\{4\}$, is measured at high enough multiplicities that one can ignore the impact of the hydrodynamization surface. These remarks bring one to conclude the following scenarios: First, the anisotropic flow in p--p collisions can be explained by conventional hydrodynamics, but there is a missing piece in the initial state models that prevents the model from producing a large negative kurtosis for ellipticity fluctuation. Second, conventional hydrodynamics alone cannot explain p--p collision observations while the initial states have the correct fluctuating properties. One candidate for the second scenario could be the presence of hydrodynamic fluctuations \cite{Kapusta:2011gt}. In particular, in the context of Gubser flow, it has been demonstrated that 
 hydrodynamic fluctuations have non-negligible effects on flow measurements of p--p collisions \cite{Yan:2015lfa}. This scenario is worth a more careful study in the future.

\section*{ACKNOWLEDGMENT}
I thank A.~Bilandzic for comments and discussions. I also thank C.~Mordasini and M.~Lesch for comments. I thank U.~A.~Wiedemann, A.~Kurkela, and W. van der Schee for discussions.  This project has received funding from the European Research Council (ERC) under the European UnionsHorizon 2020 research and innovation program (Grant No. 759257).

\appendix

\section{A simple hydrodynamic model based on the Gubser solution ({\tt GubsHyd})}\label{appA}

The initiation and hydrodynamic evolution in the {\tt GubsHyd} model are based on the Gubser solution. Although the solution is briefly mentioned in the main text, I present it with slightly more details as it is necessary for the rest of the appendix.

Instead of $\mathbb{R}^{1,3}$ space,
one is able to employ the conformal symmetry to represent a generic field $\Phi$ in $d\hat{s}^2=\Omega^{-2} ds^2$ space with $\Omega=\tau$. I show $\Phi$ in the rescaled space $d\hat{s}^2$ with a hat. Depending on the mass dimension and the type of the tensor $\Phi$, the appropriate scaling should be employed as well.  For instance, a scalar quantity $\phi$ with mass dimension $\Delta$ is written in $d\hat{s}^2$ space as $\hat{\phi}=\tau^\Delta\phi$. In this space, the Gubser solution with an elliptic perturbation on top of that is given by  \cite{Gubser:2010ze,Gubser:2010ui,Staig:2011wj}
\begin{subequations}\label{perturbedQuantitiesII}
	\begin{align}
	&\hat{\epsilon}=\frac{\hat{\epsilon}_0}{\cosh^{8/3}\rho}\left[1+4\lambda \,\delta_2(\rho) \,y(\theta,\phi) \right]+\mathcal{O}(\lambda^2),\label{21a}\\
	&\hat{u}_\mu=(-1,\;\lambda\,\delta \hat{u}_\theta,\;\lambda\,\delta \hat{u}_\phi,0)+\mathcal{O}(\lambda^2),
	\end{align}
\end{subequations}
where 
\begin{subequations}\label{ylm}
	\begin{align}
	&y(\theta,\phi)=-\sqrt{\frac{3}{8}}Y_{2,2}+\frac{1}{2}Y_{2,0}-\sqrt{\frac{3}{8}}Y_{2,-2},\\
	&\delta\hat{u}_i=\tilde{\delta}_2(\rho)\;\partial_i y(\theta,\phi)+\mathcal{O}(\lambda),\quad i=\theta,\phi,
	\end{align}
\end{subequations}
and $Y_{l,m}\equiv Y_{l,m}(\theta,\phi)$  is the spherical harmonic function. The hydrodynamic equations are given by $\partial_\mu T^{\mu\nu}=0$ where
\bea\label{a3}
T^{\mu\nu}=(p+\epsilon)u^{\mu}u^\nu+p\,\eta^{\mu\nu},
\eea
is the ideal stress-energy tensor and $\eta^{\mu\nu}=\text{diag}(-1,1,1,1)$ is the metric.  Substituting the ansatz in Eq.~\eqref{perturbedQuantitiesII} into hydrodynamic equations and using conformal e.o.s. $\epsilon=3p$,
 one obtains the following system of differential equations for $\delta_2(\rho)$ and $\tilde{\delta}_2(\rho)$:
\bea
\frac{ d}{d \rho} \begin{pmatrix}
	\delta_2(\rho) \\
	\tilde{\delta}_2(\rho)
\end{pmatrix}=-\begin{pmatrix}
	0 & -2\sech^2\rho \\
	1 & -\frac{2}{3}\tanh\rho
\end{pmatrix} \begin{pmatrix}
	\delta_2(\rho) \\
	\tilde{\delta}_2(\rho)
\end{pmatrix}.
\eea
Initial isotropic in fluid velocity assumption fixes the initial condition as
\bea
\delta_2(\rho_\hyd)=1,\qquad \tilde{\delta}_2(\rho_\hyd)=0.
\eea
This solution can be translated to hyperbolic-cylindrical coordinates via 
\bea
\tau=\frac{1}{q}\frac{\sech\rho}{\cos\theta-\tanh\rho},\qquad r=\frac{1}{q}\frac{\sin\theta}{\cos\theta-\tanh\rho}.
\eea
The connection between $(q,\hat{\epsilon}_0,\lambda,\rho_\hyd)$ and standard parameters  $(r_\rms,\epsilon_\tot,\epsilon_2,\tau_\hyd)$ has been discussed in the main text.
For the next step, I compute the hadronization of the above solution. 

I focus on events with size $r_\crit<r_\rms < R_\crit$ in which a hydrodynamized matter, ``QGP'', is formed at least in a small region in the energy density (core) while at the tail of the
energy density (corona), the system is
in the hadronic phase. The spectrum of such a system can be written as ($d\boldsymbol{p}\equiv d^3p/E$)
\bea\label{CoreCoronea}
\frac{dN_\text{QGP}}{d\boldsymbol{p}}=\frac{dN_{\text{core}}}{d\boldsymbol{p}}+\frac{dN_\text{corona}}{d\boldsymbol{p}}.
\eea 
I assume the core part evolves with hydrodynamic equations, while free streaming starts immediately after the initiation for the corona part.

I employ the Cooper-Frye prescription [see Eq.~\eqref{cff}] to obtain the particle distribution ($g=2(N_c^2-1)=16$ has been chosen).  In this paper, I have considered Boltzmann equilibrium distribution instead of a more sophisticated Bose-Einstein/Fermi-Dirac near-equilibrium distribution. 

For the core part, the particles are emitted after the hydrodynamic evolution from freeze-out surface $\Sigma^\mu=(\rho,\theta_\fo(\rho,\phi),\phi,\eta)$ with surface element
\bea
d\Sigma_\mu=-\cosh^2\rho\sin\theta_\fo(\frac{\partial\theta_\fo}{\partial \rho},-1,\frac{\partial\theta_\fo}{\partial \phi},0)d\rho \,d\phi \,d\eta.\nn\\
\eea
The functionality of $\cos\theta_\fo(\rho,\phi)$ for the Gubser background solution ($\lambda=0$) has been mentioned in Eq.~\eqref{freezeSurface}. To obtain anisotropy in the final particle distribution, however, one needs to keep the elliptic perturbation. This modifies the Eq.~\eqref{freezeSurface} to the following:
\begin{equation}\label{freezeSurfaceP}
\begin{split}
\cos\theta_\fo&(\rho,\phi)=\tanh\rho+\\
&\frac{1}{q}\left(\frac{\epsilon_\fo}{\hat{\epsilon_0}}\right)^{1/4}\sech^{1/3}\rho\left[1+\lambda \delta_2(\rho) f(\rho,\phi)\right],
\end{split}
\end{equation}
where
\begin{equation}
\begin{split}
f(\rho,\phi)=\frac{1}{8 }\sqrt{\frac{5}{\pi}}\left[1+3\cos2\phi-6\cos^2\phi\,\cos^2\theta_b(\rho)\right],
\end{split}
\end{equation}
and $\theta_b(\rho)=\cos\theta_\fo(\rho,\phi)|_{\lambda=0}$.

I start with computing the final multiplicity. Because the Gubser solution has a more straightforward analytical form in rescaled space $d\hat{s}^2$, I first calculate Eq.~\eqref{cff} in the rescaled space and then I transform $dN_{\text{core}}/d\boldsymbol{\hat{p}}$ into $dN_{\text{core}}/d\boldsymbol{p}$, using the fact  that the latter has mass dimension -2 (recall $d\boldsymbol{p}\equiv d^3p/E$):
\bea\label{tau2Measure}
\frac{dN}{d\boldsymbol{p}}=\tau^2\frac{dN}{d\boldsymbol{\hat{p}}}.
\eea 
One notes that  the four-momentum $p^\mu$ ($p_\mu p^\mu=-m^2$) in the hyperbolic-cylindrical coordinate reads as
\begin{equation} 
\begin{split}
&p^\tau=m_T\cosh(y_p-\eta),\\
&p^{r}=p_T \cos(\phi_p-\phi), \\
&p^\phi= \frac{p_T}{r} \sin(\phi_p-\phi), \\
&p^\eta=\frac{m_T}{\tau}\,\sinh(y_p-\eta),
\end{split}
\end{equation}
where $(p_T,\phi_p)$ is the transverse momentum in the polar coordinate, $m_T$ is the transverse mass, and $y_p$ is rapidity. Subsequently, one obtains the momentum in $d\hat{s}^2$ via
\bea
\hat{p}^\mu=\tau^{2}\frac{\partial \hat{x}^\mu}{\partial x^\nu} p^\nu,
\eea
where $\hat{x}^\mu=(\rho,\theta,\phi,\eta)$ and $x^{\mu}=(\tau,r,\phi,\eta)$.  
Keeping terms up to linear order in $\lambda$, the particle distribution for the core part is written as
\begin{equation}
\begin{split}\label{CooFryII}
\frac{dN_{\text{core}}}{d\boldsymbol{p}}&=\int_{\text{freeze}} \exp\left[\frac{-\hat{p}^\rho+\lambda \hat{p}^\theta\delta\hat{u}_\theta+\lambda\hat{p}^\phi\delta\hat{u}_\phi}{\tau_\fo\,T_\fo}\right],\\
&=\int_{\text{freeze}} \left[1+\lambda\left(\frac{\hat{p}^\theta\delta\hat{u}_\theta+\hat{p}^\phi\delta\hat{u}_\phi}{\tau_\fo T_\fo} \right)\right]\exp\left[\frac{-\hat{p}^\rho }{\tau_\fo\,T_\fo}\right],
\end{split}
\end{equation}
where
\begin{equation}
\begin{split}\label{CooFryInt}
\int_{\text{freeze}}&\equiv-\frac{g}{(2\pi)^3}\int_0^{2\pi}d\phi\,\int_{\rho_\hyd}^{\rho_\text{max}}\,d\rho\int_{-\infty}^\infty\, d\eta\;\\
&\hspace*{-0.2cm}\times\cosh^2\rho\sin\theta_\fo \tau_\fo^2\left(\hat{p}^\rho\frac{\partial\theta_\fo}{\partial \rho}-\hat{p}^\theta+\hat{p}^\phi\frac{\partial \theta_\fo}{\partial\phi}\right).
\end{split}
\end{equation}
The extra $\tau_\fo=\tau(\rho,\theta_\fo)$ in the measure is coming from Eq.~\eqref{tau2Measure}. The relation $\hat{T}_\fo = \tau_\fo T_\fo$ has also been used.

To calculate the multiplicity, I concentrate on $\lambda=0$ (background solution). In this case,  the integration on $\eta$ and $\phi$ in Eq.~\eqref{CooFryII} can be performed trivially. The result reads as
\begin{equation}\label{freezeOutIII}
\begin{split}
&\frac{dN_\cor}{p_T\, dp_T\,dy_p}=-\frac{g}{\pi}\int_{\rho_\hyd}^{\rho_\mmax}d\rho \\
&\hspace*{0.0cm}\times \left[A\,m_T \, I_0(\beta \, p_T)\, K_1(\alpha\, m_T) -B\,p_T \, I_1(\beta \, p_T)\, K_0(\alpha\, m_T)\right]
\end{split}
\end{equation}
where
\begin{subequations}\label{AAlpha}
	\begin{align}
	& A(\rho)=\cosh^2\rho \sin \theta_\fo \tau_\fo^4\left[\frac{\partial \theta_\fo}{\partial \rho}\left.\frac{\partial \rho}{\partial \tau}\right|_\fo-\left.\frac{\partial \theta}{\partial \tau}\right|_\fo\right],\\
	& B(\rho)=\cosh^2\rho \sin \theta_\fo \tau_\fo^4\left[\frac{\partial \theta_\fo}{\partial \rho}\left.\frac{\partial \rho}{\partial r}\right|_\fo-\left.\frac{\partial \theta}{\partial r}\right|_\fo\right],\\	
	& \alpha(\rho)=\frac{\tau_\fo}{T_\fo}\left.\frac{\partial \rho}{\partial \tau}\right|_\fo,\\
	& \beta(\rho)=\frac{\tau_\fo}{T_\fo}\left.\frac{\partial \rho}{\partial r}\right|_\fo.
	\end{align}
\end{subequations}

The particle distribution of the corona part can also be obtained by the Cooper-Frye formula, where the freeze-out surface is coincident with the hydrodynamization surface, $\Sigma^\mu=(\rho_\hyd,\theta,\phi,\eta)$:
\begin{equation}
\begin{split}
&\left.\frac{dN_\coron}{p_T dp_Td\phi_pdy_p}\right|_{\rho_\hyd}= -\frac{g}{(2\pi)^3}\int_0^{2\pi}d\phi\,\int_{-\infty}^\infty\, d\eta\,\int_{\theta=\pi}^{\theta(\rho_\hyd)} d\theta\\
& \hat{p}^\rho\cosh^2\rho_{\hyd}\sin\theta\,\tau^2\, d\theta d\phi d\eta \,e^{-\hat{p}^\rho/\hat{T}(\rho_\hyd)}
\end{split}
\end{equation}
where $\hat{T}(\rho_\hyd)=(\hat{\epsilon}_b/C_0)^{1/4}/\cosh^{2/3}\rho_\hyd$ is obtained from \eqref{perturbedQuantitiesII} at $\lambda=0$ and e.o.s. $\epsilon=C_0 T^4$.
Similar to the core part, one can trivially perform the integration on $\phi$ and $\eta$:
\begin{equation}\label{coronaFreez}
\begin{split}
&\left.\frac{dN_\coron}{p_T dp_Tdy_p}\right|_{\rho_\hyd}=-\frac{g}{\pi}\int_{\theta=\pi}^{\theta(\rho_\hyd)} d\theta \\ 
&\hspace*{-0.2cm}\times \left[A'\,m_T \, I_0(\beta' \, p_T)\, K_1(\alpha'\, m_T) -B\,p_T \, I_1(\beta' \, p_T)\, K_0(\alpha'\, m_T)\right]
\end{split}
\end{equation}
where
\begin{subequations}\label{AlphaP}
	\begin{align}
	& A'(\theta)=\cosh^2\rho_\hyd \sin \theta \,\tau^4(\rho_\hyd,\theta)\left.\frac{\partial \rho}{\partial \tau}\right|_{\rho_\hyd},\\
	& B'(\theta)=\cosh^2\rho_\hyd \sin \theta \tau^4(\rho_\hyd,\theta)\left.\frac{\partial \rho}{\partial r}\right|_{\rho_\hyd},\\	
	& \alpha'(\theta)=\frac{\tau^2(\rho_\hyd,\theta)}{\hat{T}(\rho_\hyd)}\left.\frac{\partial \rho}{\partial \tau}\right|_{\rho_\hyd},\\
	& \beta'(\theta)=\frac{\tau^2(\rho_\hyd,\theta)}{\hat{T}(\rho_\hyd)}\left.\frac{\partial \rho}{\partial r}\right|_{\rho_\hyd}.
	\end{align}
\end{subequations}
The integrals in Eqs.~\eqref{freezeOutIII} and \eqref{coronaFreez} can be computed numerically.
Referring to Eq.~\eqref{CoreCoronea}, one obtains the total multiplicity, $dN_\qgp/d\boldsymbol{p}|_{\lambda=0}$, by adding the outcome of two integrals. I have prepared the integrals in a Mathematica package where the $p_T$ dependent and $p_T$ integrated multiplicity in the unit rapidity are available:
\be
\begin{aligned}
	&\scriptsize\verb||\normalsize \verb| dNtotalOverPdPdyp[|  \epsilon_\tot \verb|,| r_\rms\verb|,| p_T\verb|]|,\\
	&\scriptsize\verb||\normalsize \verb| dNtotalOverdyp[|  \epsilon_\tot\verb|,| r_\rms\verb|,| p_{T,\text{min}}\verb|,| p_{T,\text{max}}\verb|]|.
\end{aligned}
\ee
The total multiplicity in the unit rapidity $n_\tot(r_{\text{rms}},\epsilon_\text{tot})$ is obtained from the second function with $p_T$ range from zero to infinity. 
One can also change the parameters of final particle distribution from $y_p$ to space-time pseudorapidity via 
\begin{equation}
\begin{split}
\frac{dN}{d\eta}&=\sqrt{1-\frac{m^2}{m_T^2\cosh^2y(\eta)}}\,\frac{dN}{dy_p},\\
y(\eta)&=\frac{1}{2}\log\left(\frac{\sqrt{p_T^2\cosh^2\eta+m^2}+p_T\sinh\eta}{\sqrt{p_T^2\cosh^2\eta+m^2}-p_T\sinh\eta}\right).
\end{split}
\end{equation}
Now by using the above and by inverting $n_\tot(r_{\text{rms}},\epsilon_\text{tot})$ to find  $\epsilon_\tot(n_\tot,r_\rms)$, one obtains $dN_\qgp/p_Tdp_Td\eta|_{\lambda=0}$ and $N_\tot|_{\lambda=0}$ as a function of total multiplicity in the unit rapidity, rms radius, transverse momentum, and space-time pseudorapidity: 
\be
\begin{aligned}\label{A22}
	&\scriptsize\verb||\normalsize \verb| dNtotalOverPdPdEta[| n_\tot\verb|,| r_\rms\verb|,| p_T\verb|,| \eta\verb|]|,\\
	&\scriptsize\verb||\normalsize \verb| Ntotal[| n_\tot\verb|,| r_\rms\verb|,| p_{T,\text{min}}\verb|,| p_{T,\text{max}}\verb|,| \eta_{\text{min}}\verb|,| \eta_{\text{max}}\verb|]|.
\end{aligned}
\ee

Having found the energy dependence of multiplicity, one can translate the lower and upper bound of the system size into multiplicity. The total energy density for a system at critical size can be obtained from Eq.~\eqref{theInequalityI},
\bea
\epsilon_\crit=\frac{4\pi}{3}\left(\tau_\hyd^4\epsilon_\fo\right)\frac{1}{r_\crit^2}.
\eea
Substituting $\epsilon_\crit$ into $n_\tot(r_\rms,\epsilon_\tot)$, one obtains a critical value for the multiplicity. Another critical value can be found by inverting $R_\crit(\epsilon_\tot)$ (see Eq.~\eqref{upperbound}) and substituting the result into $n_\tot(r_\rms,\epsilon_\tot)$. Both of these critical multiplicities are available in the package,
\be
\begin{aligned}
	&\scriptsize\verb||\normalsize \verb| ncrit[| r_\rms\verb|,| p_{T,\text{min}}\verb|,| p_{T,\text{max}}\verb|]|,\\
	&\scriptsize\verb||\normalsize \verb| Ncrit[| r_\rms\verb|,| p_{T,\text{min}}\verb|,| p_{T,\text{max}}\verb|]|.
\end{aligned}
\ee
The above functions are shown with thick black curves in Fig.~\eqref{k2} for $0.3<p_T<3\,\text{GeV}$.
In massless limit $m=0$, both $n_\tot(r_\rms,\epsilon_\tot)$ and $n_\crit
$ (I assume a full range $p_T$ integration) can be obtained analytically,
\begin{equation}
\begin{split}
n_\tot(r_\rms,\epsilon_\tot)&=\frac{g}{C_0^{3/4}}\sqrt{\frac{3}{\pi^3}}\left[\tau_\hyd \,\epsilon_\fo^{1/4}\,r_\rms\, \epsilon_\tot^{1/2}\right],\\
n_\crit&=\frac{4g}{\pi C_0^{3/4}}\left[\tau_\hyd^3 \,\epsilon_\fo^{3/4}\right].
\end{split}
\end{equation}
As can be seen in the massless limit, $n_\crit$ has no $r_\rms$ dependence.

Now, one can compute $k_2(r_\rms,n_\tot)=\left.(\partial v_2 /\partial \epsilon_2)\right|_{\epsilon_2=0}$. The final particle distribution is obtained via
\begin{equation}\label{6}
\begin{split}
\frac{d N_\qgp}{d\boldsymbol{p}}\approx\left.\frac{d N_\qgp}{d\boldsymbol{p}}\right|_{\lambda=0}+\left. \frac{d}{d\lambda}\frac{d N_\cor}{d\boldsymbol{p}}\right|_{\lambda=0}\lambda.
\end{split}
\end{equation}
The first term in the right-hand side is already computed. In the second term, only the core part contributes to momentum anisotropy at the final state. This is because I have assumed free streaming for the corona part. Therefore, the elliptic flow can be obtained by
\begin{equation}\label{v2Definition}
\begin{split}
v_2&=\left(\left.\frac{d N_\qgp}{d\boldsymbol{p}}\right|_{\lambda=0}\right)^{-1}\\
&\hspace*{1cm}\times\left[\int d\phi_p \cos2\phi_p\left(\left.\frac{d}{d\lambda}\frac{d N_\cor}{d\boldsymbol{p}}\right|_{\lambda=0}\right)\right]\lambda.
\end{split}
\end{equation}
One can write the linearized $dN_\cor/d\boldsymbol{p}$ in Eq.~\eqref{CooFryII} with respect to $\lambda$ as follows:
\begin{equation}\label{38}
\begin{split}
\frac{dN_\cor}{d\boldsymbol{p}}&= \frac{g}{(2\pi)^3}\int d\phi d\eta \int_{\rho_\hyd(\lambda)}^{\rho_\mmax(\lambda)} d\rho\\
&\hspace*{1cm}\times\left[\Q_0 e^{\chi_0} +\lambda \left(\Q_1+\Q_0 \chi_1\right) e^{\chi_0}\right],
\end{split}
\end{equation}
where
\begin{equation}
\begin{split}
&\Q\equiv-\cosh^2\rho\sin\theta_\fo \tau_\fo^2\left(\hat{p}^\rho\frac{\partial\theta_\fo}{\partial \rho}-\hat{p}^\theta+\hat{p}^\phi\frac{\partial \theta_\fo}{\partial\phi}\right),\\
&\chi\equiv \frac{1 }{\tau_\fo\,T_\fo}\left[-\hat{p}^\rho+\lambda\hat{p}^\theta\delta\hat{u}_\theta+\lambda\hat{p}^\phi\delta\hat{u}_\phi\right].
\end{split}
\end{equation}
and
\begin{equation}
\begin{split}
&\Q = \Q_0 +\lambda \Q_1 +\mathcal{O}(\lambda^2),\\
&\chi = \chi_0 +\lambda \chi_1 +\mathcal{O}(\lambda^2).
\end{split}
\end{equation}
For a given function $f(x,t)$, I employ the following calculus identity,
\begin{equation}
\begin{split}
\frac{d}{dt}\int_{a(t)}^{b(t)}&f(x,t)dx=\int_{a(t)}^{b(t)} \frac{\partial f}{\partial t}dx\\
&+\frac{\partial b(t)}{\partial t}f(b(t),t)-\frac{\partial a(t)}{\partial t}f(a(t),t).
\end{split}
\end{equation}
By using the above identity, the derivative of Eq.~\eqref{38} at $\lambda=0$ is given by
\begin{equation}\label{41}
\begin{split}
&\frac{d}{d\lambda}\frac{dN_\cor}{d\boldsymbol{p}}\Bigg|_{\lambda=0}=\frac{g}{(2\pi)^3}\int d\phi d\eta\\
&\hspace*{1cm}\Bigg[\left(\int_{\rho_\hyd(0)}^{\rho_\mmax(0)} d\rho\;e^{\chi_0}\left(\Q_1+\Q_0 \chi_1\right)\right)\\
&(\Q_0 e^{\chi_0}|_{\rho_\mmax(0)})\rho'_\mmax(0)-(\Q_0 e^{\chi_0}|_{\rho_\hyd(0)})\rho'_\hyd(0)\Bigg],
\end{split}
\end{equation}
where the prime in $\rho'_\hyd(\lambda)$ is the derivative with respect to $\lambda$.
One notes that, by definition, $\cos\theta_\fo(\rho_\mmax)=1$, which immediately leads to $\sin\theta_\fo(\rho_\mmax)=0$. Given that there is a  term containing $\sin\theta_\fo$ in $\Q_0$, the  boundary term $\Q_0 e^{\chi_0}|_{\rho_\mmax(0)}$ in Eq.~\eqref{41} is identically vanishing. The second boundary term, however, is not zero.
To find $\rho'_\hyd(0)$, one needs to compute $\rho_{\hyd}$ at finite $\lambda$. The quantity $\hat{\epsilon}_0$ up to the linear term in $\lambda$ is given by
\bea\label{42}
\hat{\epsilon}_0=\frac{3\,\epsilon_\tot \,r_\rms^2}{4\pi\cosh^{4/3}\rho_\hyd}\left(1+\frac{\lambda}{\sqrt{5\pi}}\right)+\mathcal{O}(\lambda^2).
\eea
Noting $\epsilon=\hat{\epsilon}/\tau^4$ and substituting Eq.~\eqref{42} into Eq.~\eqref{21a}, one obtains
\begin{equation}\label{43}
\begin{split}
\cosh^4\rho_\hyd&(\lambda)=\cosh^4\rho_\hyd(0)\\
&\left[1+\left(\frac{1}{\sqrt{5\pi}}+4y(\theta_\fo,\phi)\right)\lambda\right]+\mathcal{O}(\lambda^2),
\end{split}
\end{equation}
where in the above $\theta_\fo\equiv\theta_\fo(\rho_\hyd(0),\phi) $. One can compute $\rho'_\hyd(0)$ from Eq.~\eqref{43}.

In Eq.~\eqref{41}, the integration over $\phi$ and $\eta$ is tedious but analytically doable, while the integration over $\rho$ needs to be done numerically. Symbolically, one can write Eq.~\eqref{41} as follows:
\begin{equation}
\begin{split}
&\left.\frac{d}{d\lambda}\frac{dN_\cor}{d\boldsymbol{p}}\right|_{\lambda=0}=\frac{g}{2\pi}\left[\V_0(p_T)+2\V_2(p_T)\cos2\phi_p \right],
\end{split}
\end{equation}
where $\V_0(p_T)$ and $\V_2(p_T)$ are two complicated functions depending on the standard quantities including a numerical integration over $\rho$. 
Using Eq.~\eqref{v2Definition}, one finds 
\begin{equation}
\begin{split}
k_2(\epsilon_\tot&,r_\rms,p_T)=\left.\frac{\partial v_2}{\partial \epsilon_2}\right|_{\epsilon_2=0}=\\
&\frac{\sqrt{5\pi}}{3}\left(\left.\frac{d N_\qgp}{d\boldsymbol{p}}\right|_{\lambda=0}\right)^{-1}\,\V_2(p_T).
\end{split}
\end{equation}
where I have substituted $\lambda=\left(\sqrt{5\pi}/3\right)\epsilon_2$. This function and its $p_T$ integrated version are available in the Mathematica package,
\be
\begin{aligned}\label{50}
	&\scriptsize\verb||\normalsize \verb| k2Energy[| \epsilon_\tot\verb|,| r_\rms\verb|,| p_{T}\verb|]|,\\
	&\scriptsize\verb||\normalsize \verb| k2Energy[| \epsilon_\tot\verb|,| r_\rms\verb|,| p_{T,\text{min}}\verb|,| p_{T,\text{max}}\verb|]|.
\end{aligned}
\ee
Also, by substituting $\epsilon_\tot(n_\tot,r_\rms)$ into $k_2(\epsilon_\tot,r_\rms)$, I obtain hydrodynamic response in terms of rms radius and multiplicity in the unit rapidity:
\be
\begin{aligned}\label{a38}
	&\scriptsize\verb||\normalsize \verb| k2[| n_\tot\verb|,| r_\rms\verb|,| p_{T}\verb|]|,\\
	&\scriptsize\verb||\normalsize \verb| k2[| n_\tot\verb|,| r_\rms\verb|,| p_{T,\text{min}}\verb|,| p_{T,\text{max}}\verb|]|.
\end{aligned}
\ee
The second function in the above within range $0.3<p_T<3\,\text{GeV/c}$ has been employed to plot Fig.~\eqref{k2}.

\section{Validation of {\tt GubsHyd} via {\tt  iEBE-VISHNU}}\label{validationApp}

\begin{figure}[t]
	\begin{center}
		\includegraphics[scale=0.63]{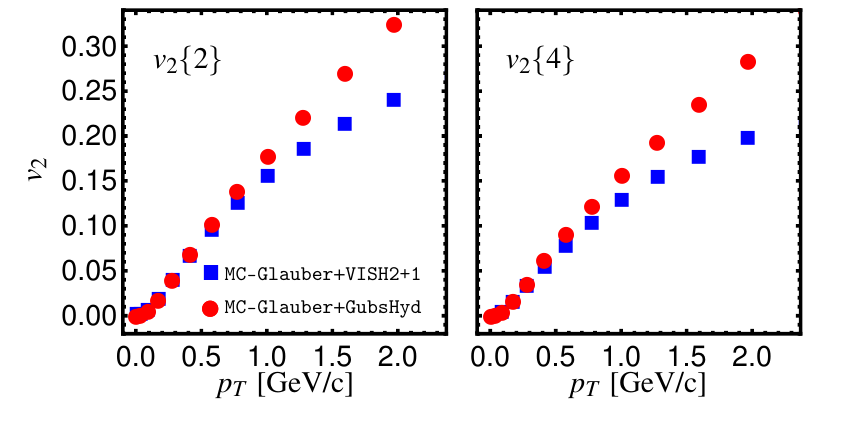}
		\caption{Differential anisotropic flow $v_2\{2\}$ and $v_2\{4\}$ from {\tt MC-Glauber}+{\tt VISH2+1} and {\tt MC-Glauber}+{\tt GubsHyd}.} 
		\label{vnpt}
	\end{center}
\end{figure}

The dissipative effects in both hydrodynamic evolution and freeze-out have been ignored in the {\tt GubsHyd} model.
In the main text, I assume that the contribution of dissipative effects on integrated $v_2\{2\}$ and $v_2\{4\}$ can be corrected via an overall factor $\chi$ (see Eq.~\eqref{37}). I have examined this assumption by comparing the prediction of {\tt MC-Glauber}+{\tt GubsHyd} and {\tt MC-Glauber}+{\tt VISH2+1} ({\tt iEBE-VISHNU}) with nonvanishing shear viscosity over entropy density for integrated $v_2\{2\}$ and $v_2\{4\}$ of charged pions (see Fig.~\ref{GubsHydVISHNUcomp}). In this appendix, I extend the comparison to the following observables: 
$p_T$ spectrum, differential $v_2\{2\}$, and $v_2\{4\}$ and centrality dependence of the multiplicity for charged pions. These comparisons lead to a clearer picture of the accuracy of {\tt GubsHyd} and the limitations that  should be considered in the conclusions (see Ref.~\cite{Dusling:2007gi} for a comprehensive study about the role of the dissipative effects on anisotropic flow and $p_T$ spectrum).

  \begin{figure}[t]
	\begin{center}
		\includegraphics[scale=0.62]{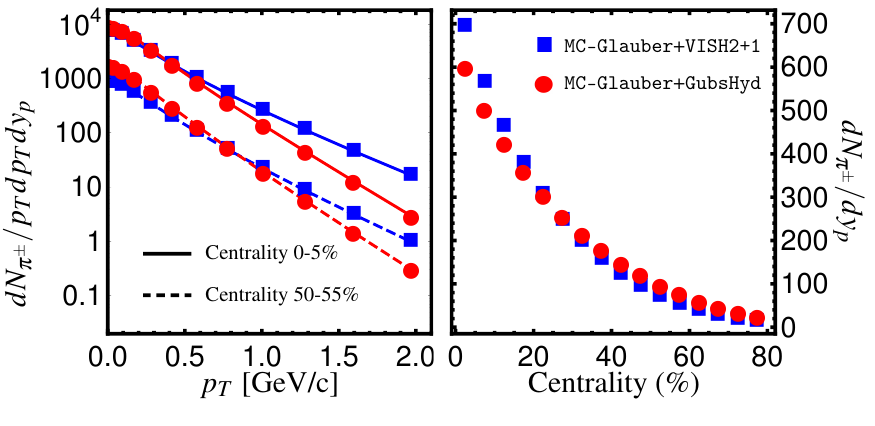}
		\caption{Charged pions $p_T$ spectrum and centrality dependence of the charged pions multiplicity from {\tt MC-Glauber}+{\tt VISH2+1} and {\tt MC-Glauber}+{\tt GubsHyd}.} 
		\label{spec}
	\end{center}
\end{figure}

The contribution of dissipative effects on elliptic flow can be explained as follows: In the course of evolution, compared to an ideal fluid, the initial elliptic shape of the energy density becomes more round at the presence of the dissipative effects. Consequently, $v_2$ fluctuates in a narrower range event by event.    This argument explains that $\chi$ in Eq.~\eqref{37} should be smaller than unity to get a reasonable agreement between {\tt GubsHyd} and {\tt iEBE-VISHNU} in Fig.~\ref{GubsHydVISHNUcomp}. One notes that the dissipative effects are dynamical and cannot be fully corrected via a simple scaling factor. In Fig.~\ref{vnpt}, the $p_T$ dependence of $v_2\{2\}$ and $v_2\{4\}$ is shown where parameters are tuned as those mentioned in Sec.~\ref{secIII}. As seen, the {\tt GubsHyd} prediction starts deviating from {\tt iEBE-VISHNU} prediction above $p_T\sim 1$GeV/c, meaning the dissipative effects have more influence on the harder particles.
Since the particles with $p_T\lesssim 1$GeV/c are more abandoned, the part with  $p_T\gtrsim 1$GeV/c is less important in integrated $v_2\{2k\}$ observables. As a result, a constant $\chi$ leads to a reasonable approximation for integrated $v_2\{2\}$ and $v_2\{4\}$.

 Now, I discuss the $p_T$ spectrum and multiplicity of charged pions.  The dissipative effects lead to entropy production and consequently increase the number of particles in the final state. As a result, one expects the following relation:
 \bea\
 \frac{dN^{\text{\tiny Gub}}}{d\boldsymbol{p}}=\chi'\frac{dN^{\text{\tiny IdealGub}}}{d\boldsymbol{p}},
 \eea 
 where $\chi'$ should be greater than unity. The numerical value of $\chi'$ has no impact on the observables I  discuss in the main text, but I analyze it here for completeness. Although the dissipation effects act dynamically on multiplicity and cannot be fully corrected by a constant $\chi'$, it is still a good approximation to assume it is a constant in a wide range of centralities, similar to integrated $v_2\{2\}$ and $v_2\{4\}$.  By demanding that the average of the number of charged pions in the unit rapidity, $\la dN_{\pi^\pm}/dy_p\ra$, is the same in the centrality range 0$-$80\% for  {\tt GubsHyd} and {\tt iEBE-VISHNU}, one finds $\chi'\approx 1.4$. Using this number, I have plotted   the $p_T$ spectrum of charged pions in Fig.~\ref{spec}~(left). As seen from the figure, the $p_T$ spectrum of {\tt iEBE-VISHNU} has a smaller slope compared to {\tt GubsHyd}. The reason is that, similar to differential $v_2$, the dissipative effects are more significant at higher $p_T$. As a result, there is more entropy production and more particles at higher $p_T$.  The centrality dependence of charged pion multiplicity is depicted in Fig.~\ref{spec}~(right).  One observes that the overall trends of {\tt GubsHyd} and {\tt iEBE-VISHNU} are in a reasonable agreement even though the prediction from {\tt iEBE-VISHNU} shows a steeper trend.

\end{document}